\documentstyle[12pt,epsfig]{article}

\def\beq{\begin{equation}} \def\eeq{\end{equation}}
\def\beqa{\begin{eqnarray}} \def\eeqa{\end{eqnarray}}
\def\bce{\begin{center}}  \def\ece{\end{center}}
\def\bfig{\begin{figure}}  \def\efig{\end{figure}}
\def\bit{\begin{itemize}}    \def\eit{\end{itemize}}
\def\ben{\begin{enumerate}}    \def\een{\end{enumerate}}
\def\npb{Nucl. Phys. {\bf B}} \def\plb{Phys. Lett. {\bf B}}
\def\prd{Phys. Rev. {\bf D}}
\def\prp{Phys. Rep.} \def\prl{Phys. Rev. Lett.}

\begin{document}

\rightline{CERN-TH/98-357}
\rightline{FNT/T-98/12}

\vspace{1.truecm}

\begin{center}
{\Large \bf Six-fermion production and Higgs boson\\}
{\Large \bf physics at future $e^+ e^-$ colliders}
\end{center}

\noindent
\begin{center}
Fabrizio GANGEMI$^{1,2}$, Guido MONTAGNA$^{1,2}$, Mauro MORETTI$^{3,4}$, \\
Oreste NICROSINI$^{2,1}$  and Fulvio PICCININI$^{2,1}$
\end{center}

\vskip 0.3cm
\noindent
\begin{center}
$^1$ Dipartimento di Fisica Nucleare e Teorica, Universit\`a di Pavia, Italy\\
$^2$ Istituto Nazionale di Fisica Nucleare, Sezione di Pavia, Italy\\
$^3$ CERN - Theory Division, Geneva, Switzerland\\
$^4$ Dipartimento di Fisica, Universit\`a di Ferrara and INFN, Sezione di Ferrara, 
Italy
\end{center}

\vskip 0.5cm

\begin{abstract}

\vskip 0.2cm
The six-fermion production processes $e^+e^-\rightarrow 
q\bar q l^+ l^-\nu\bar\nu$, with all the lepton flavours
and $q=u,d,c,s$, relevant to the study of the 
intermediate-mass Higgs boson at future $e^+e^-$ linear colliders,
are analysed. A Monte Carlo program, taking into account the whole set of 
tree-level scattering amplitudes and the relevant radiative effects,
is developed to provide integrated cross sections and generation of unweighted
events. The complete calculation is compared with the available
results of real Higgs production, and the opportunities of
precision studies with event generation are discussed, 
demonstrating the relevance of a full six-fermion calculation. 
Numerical results for integrated cross sections with various kinematical
cuts and including radiative effects are given and commented.
In the analysis of event samples, several distributions are studied
and found to be sensitive to the presence and to the properties of the
Higgs boson.

\end{abstract}

\vspace{0.5truecm}
\begin{small}
\end{small}

\section{Introduction}
The search for the Higgs boson and the study of its properties will be among
the most important tasks of elementary particle physics at future $e^+e^-$
linear colliders at the TeV scale (NLC)~\cite{lc}.

The research  carried on at present colliders can explore an interval of Higgs
masses below $\sim 100$~GeV at LEP2~\cite{mcn} or possibly $120$--$130$ GeV at the
upgraded Tevatron~\cite{tevatron}. The remaining mass range, up to
$\sim 800$ GeV, will be in the reach of the future colliders LHC~\cite{LHC} and
NLC. In particular the precision studies that will be possible in the clean
environment of NLC will be of great help in the determination of the Higgs
boson properties.

A range of particular interest for the Higgs mass is between $100$ and
$200$ GeV, as many arguments both of experimental and of theoretical nature
indicate. Indeed a lower limit of $\sim 90$ GeV is given by recent results
in the direct search at LEP2~\cite{mcn}, while from fits to electroweak
precision data an upper limit of $\sim 280$ GeV at $95\%$ C.L. is
obtained~\cite{karlen}.

In this range two mass intervals may be considered: for $m_H\leq 140$ GeV the
Higgs decays mainly into $b\overline b$ pairs, while for $m_H\geq 140$ GeV the
decays into $WW$ and $ZZ$ pairs become dominant.
Therefore in the first case the mechanisms of Higgs production relevant to 
$e^+e^-$ colliders, Higgs-strahlung and $VV(V=W,Z)$ fusion, give rise to
signatures that contain four fermions in the final state, which have been
extensively studied in the recent past~\cite{lep2}--\cite{dpeg}.
In the second case, in which $m_H\geq 140$ GeV, six fermions are produced in the
final state.
More generally, six-fermion ($6f$) final states come from other relevant
processes at NLC, such as $t\bar t$ and three-gauge-boson 
production~\cite{keystone}.

As shown by complete four-fermion calculations for $WW$ and light Higgs physics
at LEP2~\cite{lep2}--\cite{dpeg,wwwg,wweg}, full calculations of $6f$
production
processes allow one to keep phenomenologically relevant issues under control, such
as off-shellness and interference effects, background contributions and spin
correlations.
Some calculations of such processes $e^+ e^- \to 6$f  
have recently been performed~\cite{to1}--\cite{gmmnp97},  
with regard to {\it top}-quark, Higgs boson  and $WWZ$ physics at NLC. 
Moreover recent progress in the calculation of processes 
$e^+ e^- \to 6$~jets~\cite{sixj} and of $2\to$ up to $8$ partons
QCD amplitudes~\cite{alpha1,dkp}
should be mentioned for their relevance in QCD tests
at lepton and hadron machines.
 These calculations rely upon different computational techniques, such  
as helicity amplitude methods for the evaluation of the very large number  
of Feynman diagrams associated to the process under
examination, or iterative numerical algorithms, where the transition  
amplitudes are calculated numerically without using Feynman diagrams. 
 
Concerning Higgs physics, an analysis of the processes
$e^+ e^- \to \mu^+ \mu^- u \bar d \tau^- \bar \nu_\tau$ and  
$e^+ e^- \to \mu^+ \mu^- u \bar d e^- \bar \nu_e$ has been performed
 in ref.~\cite{sixfzpc},  
where the Higgs can be produced by Higgs-strahlung and  
the subsequent decay proceeds through $W^+ W^-$ pairs.  
Special attention has been devoted to the calculation of the Higgs boson  
signal and of its Standard Model background, with special emphasis on the
determination and analysis of angular correlation variables, particularly 
sensitive to the presence and to the spinless nature of the Higgs particle.
The $6f$ final states, where the Higgs signal gives  
rise to two charged currents, have also been considered in  
ref.~\cite{to1}, studying cross sections and invariant mass  
distributions for the processes  
$e^+ e^- \to f \bar f q \bar q' f' \bar f''$. 

The case of the Higgs boson decay in neutral currents has been  
briefly addressed for the signal alone in ref.~\cite{gmmnp97} with  
the study of the reaction $e^+ e^- \to e^+ e^- \nu_e \bar \nu_e  
u \bar u$. The aim of the present paper is to complete and extend  
the analysis of ref.~\cite{gmmnp97} to include general $q \bar q$  
neutral currents contributions and the effect of the contributions from 
undetectable different-flavour neutrinos, in such a way as to provide  
realistic predictions for processes 
$e^+ e^- \to l^+ l^- \nu \bar \nu q \bar q$ at the parton level.
In the following, $b$-quark tagging will be assumed, leaving aside
$b\bar b$ final states, which lead to an interplay between Higgs and
{\it top} physics and will be studied elsewhere.
Consisting of only two jets, the processes considered in the present  
paper are free from QCD backgrounds.

The outline of the paper is as follows. In Section 2 the physical process is
presented and the main technical issues of the calculation are explained.
In Section 3 several numerical results are shown and discussed and the
potentials of full $6f$ calculations are stressed. Finally,
Section 4 contains the conclusions.

\section{Physical process and computing technique}

The production of an intermediate mass Higgs boson is studied in the
process $e^+e^-\to q\overline q l^+ l^- \nu\overline\nu$, where a sum is
made over the contributions of the $u,d,c$ and $s$ quarks, of the three
neutrinos and of $l=e,\mu,\tau$.
Particular attention will be devoted to the signature
$q\overline q e^+ e^- \nu\overline\nu$.

One of the interesting features of this process is the presence of both
charged current and neutral current contributions~\cite{to1}, which is a
situation never
studied before, since all the six-fermion signatures analysed in the
literature~\cite{to1}--\cite{sixfzpc} involve only charged current
contributions. Moreover
this class of processes receives contribution from diagrams with up to three
$t$-channel gauge bosons. This feature is of particular interest
because of the large  centre-of-mass (c.m.) energy, $\sqrt s$, at which the NLC
will operate. These topologies enhance the cross section with growing $s$.
The capability to provide predictions for processes with many electrons and
electron-neutrinos in the final state is therefore crucial to discuss NLC 
physics.
The present study demonstrates the possibility to deal successfully
with the dynamics calculation and phase-space integration
of this class of final states.
Another important property is that the process is
free from QCD backgrounds because only two jets are produced.
As a drawback, the total cross section is smaller than
in the $6f$ processes with four or six jets.

However, the sums over quark, charged lepton and neutrino flavours, as well as the
combined action of different mechanisms of production (see fig.~\ref{fig:6fd}), 
contribute to give a
significant size to the cross section, so that, assuming an
integrated luminosity of 
$500$ fb$^{-1}$/yr and a Higgs mass of, say, $185$ GeV, more than $1000$
events can be expected at a c.m.
energy of $360$ GeV and more than $2000$ at
$800$ GeV (see fig.~\ref{fig:6fsscan20}). In particular, as will be seen in the
numerical results, the presence of the $t$-channel contributions of vector
boson fusion gives an enhancement of the cross sections at very high energies.

The diagrams containing a resonant Higgs boson coupled to gauge bosons for the
$q\overline q e^+ e^- \nu\overline\nu$ final state are shown in
fig.~\ref{fig:6fd}. As can be seen, there are four terms of the Higgs-strahlung
type and two of the fusion type. At relatively low energies, $\sqrt{s}\le 500$ GeV,
the process of Higgs-strahlung dominates and, in particular, the charged
current term is the most important one. As the energy is increased, the
$t$-channel terms of vector boson fusion become more and more important, as
they grow with increasing $s$, and they are dominant above $500$ GeV.
The diagrams for the processes with $\mu^+\mu^-$ or $\tau^+\tau^-$ instead of
$e^+e^-$ and/or $\bar\nu_{\mu,\tau} \nu_{\mu,\tau}$
instead of $\bar\nu_e \nu_e$
in the final state are a subset of those illustrated here.

\bce
\bfig
\bce
\epsfig{file=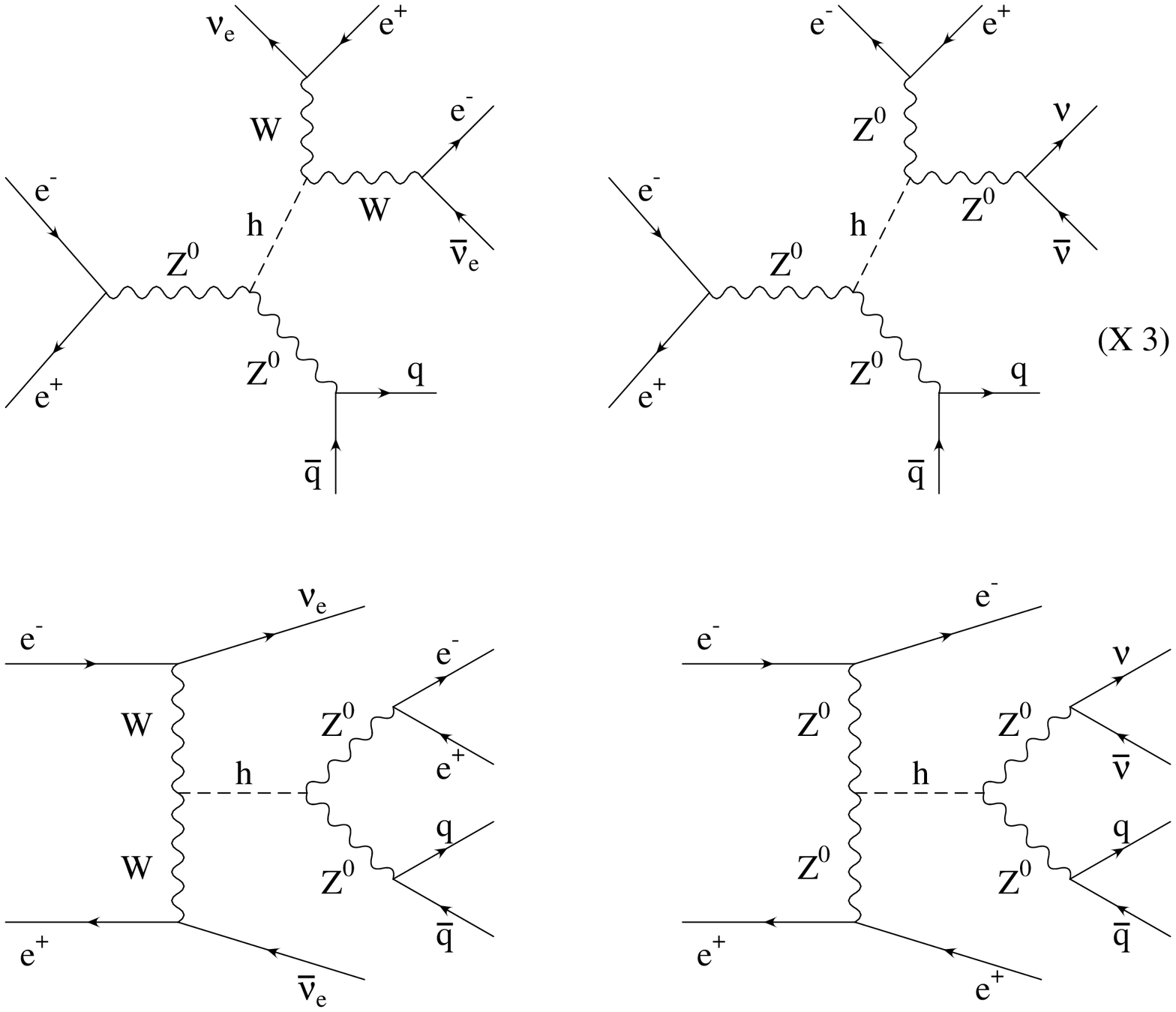,height=8.cm,width=10.cm}
\caption{\small Feynman diagrams for the process 
$e^+e^-\to q\overline q e^+ e^- \nu\overline\nu$ with a resonant Higgs boson.}
\label{fig:6fd}
\ece
\efig
\ece

The full set of diagrams containing a physical Higgs boson coupled to
gauge bosons includes also those shown in fig.~\ref{fig:6fdht}. These
contributions are non-resonant, as the Higgs is exchanged in the $t$-channel,
and their size can be expected to be negligible at low energies; at high
energies, however, they play
an important r\^ole in preserving gauge invariance and unitarity of the
$S$-matrix. This point will be discussed in more detail in the next section.
\bce
\bfig
\bce
\epsfig{file=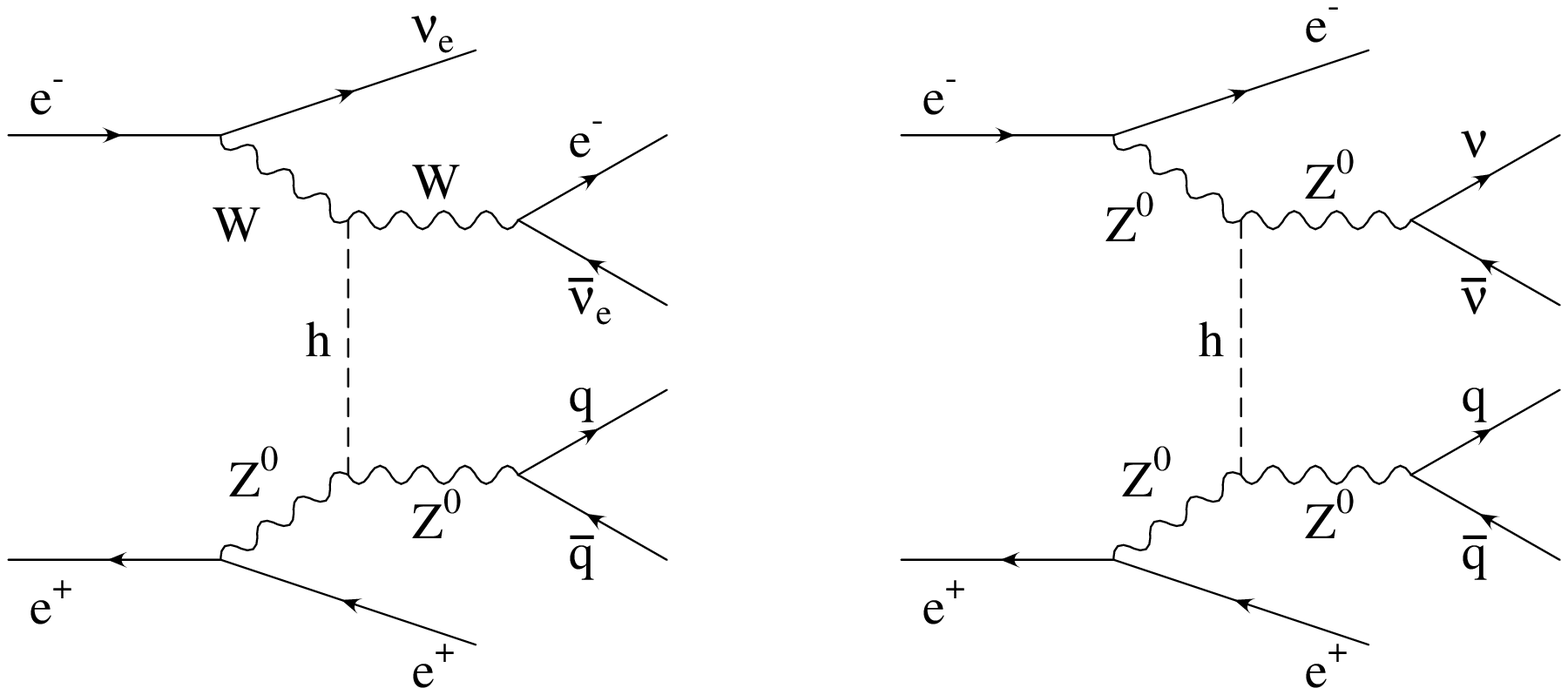,height=4.cm,width=10.cm}
\caption{\small Feynman diagrams with a non-resonant Higgs boson (two similar
diagrams are also obtained from these by particle exchanges).}
\label{fig:6fdht}
\ece
\efig
\ece

The total number of tree-level Feynman diagrams for the process under
examination is of the order of one thousand, which makes the calculations
very complicated for what concerns the determination of the transition matrix
element as well as the phase-space integration.

The transition matrix element is calculated by means of ALPHA 
~\cite{alpha}, an algorithm that allows the calculation of scattering 
amplitudes at the tree level without the use of Feynman diagrams. 
The results produced by this algorithm in a large number of calculations of
multi-particle production are in full agreement with those obtained by programs
using traditional techniques ~\cite{smwg,wweg,alpha}--\cite{mmnp}.
This fact may be considered as a significant test of ALPHA.

Anyway, also in the present work, checks have been made, reproducing by means
of the helicity amplitude method ~\cite{hel} some of the results given by
ALPHA for the Higgs ``signal'' (which is defined below) and finding perfect
agreement.

For the integration over the phase space, as was already done in
refs. \cite{sixfzpc,gmmnp97}, a code has been developed by adapting the Monte Carlo 
program HIGGSPV/WWGENPV \cite{higgspv,wwgenpv}, originally developed to treat
four-fermion production, to make $6f$ calculations.
The code can be used to perform Monte Carlo integrations
and obtain cross sections, or to generate samples of unweighted events.
Kinematical cuts can be introduced to simulate realistic experimental
conditions. The effects of initial-state radiation (ISR)~\cite{sf} and
beamstrahlung ~\cite{circe} are taken into account by means of the standard
convolution formula

\beq
  \sigma =\int dz_1dz_2D_{BS}(z_1,z_2;s)\int dx_1dx_2D(x_1,s)D(x_2,s)
  \hat\sigma (z_1,z_2;x_1,x_2;s)\ .
\eeq

An accurate importance sampling procedure is required in the Monte Carlo
integration to take care of the complicated structure of ``singularities'' in
the integrand. This structure results from the presence of several mechanisms of
Higgs production, and also from additional sources of variance among the very
large number of background diagrams present in the matrix element.

The ``singularities'' given by different terms correspond to different regions
of the (14-dimensional) phase space and in general must be treated with
different sets of integration variables. As a consequence, a multichannel
importance sampling technique is needed. If $n$ channels are introduced, the
integral is written as
\beq
\int f({\bf x})d\mu({\bf x}) = \sum_{i=1}^n\int {f({\bf x}^{(i)})\over
                                p({\bf x}^{(i)})}p_i({\bf x}^{(i)})
                                d\mu_i({\bf x}^{(i)}),
\qquad p({\bf x})=\sum_{i=1}^n p_i({\bf x}),
\eeq
where each ${\bf x}^{(i)}$ is a set of integration variables with a
corresponding measure $d\mu_i$, and $p_i$ is a suitably normalized
distribution function for the importance sampling in the $i$-th channel.

The choice of integration variables is made within two kinds of phase-space
decompositions, corresponding to two diagram topologies: $s$-channel,
based on the Higgs-strahlung terms, and $t$-channel, based on the fusion terms,
as illustrated in fig.~\ref{fig:6fcamp}.
\bce
\bfig
\bce
\epsfig{file=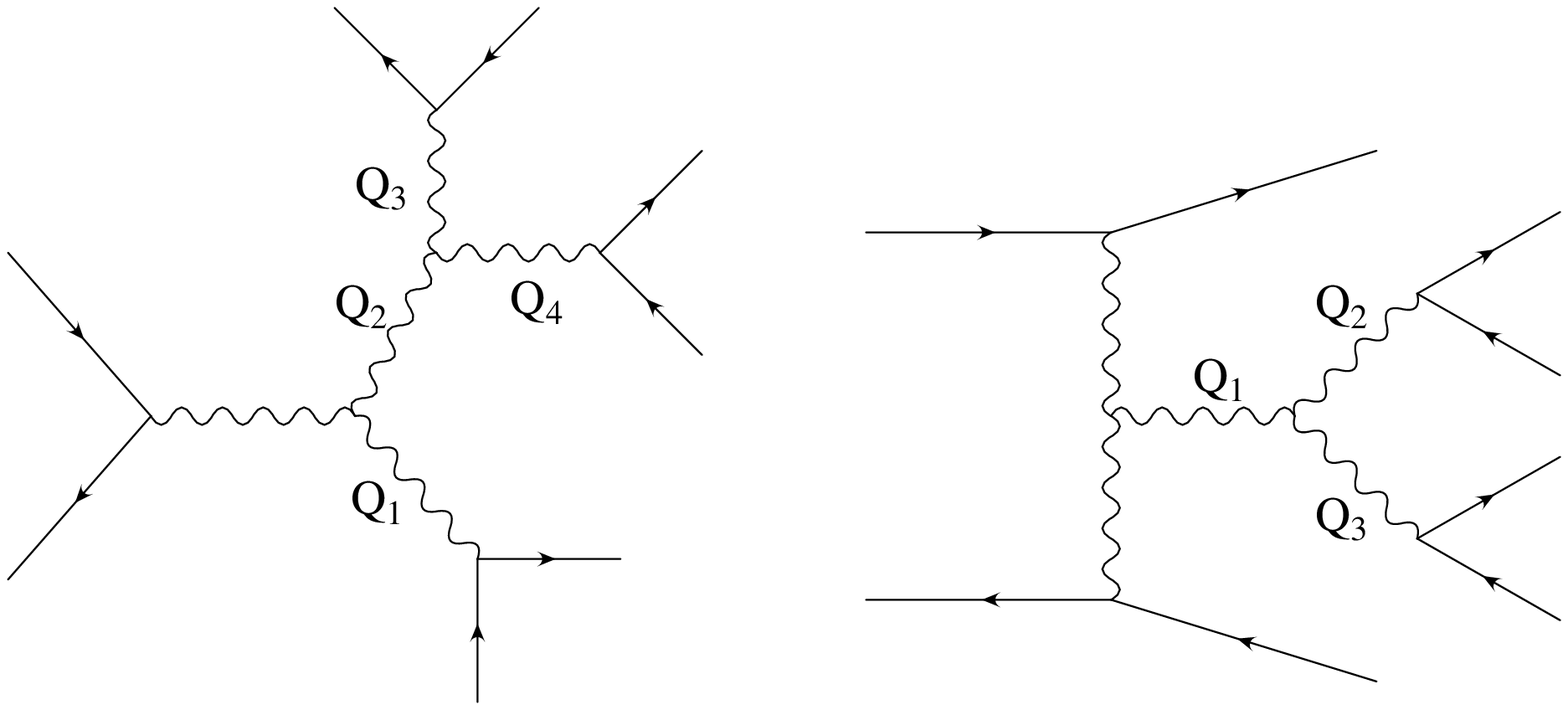,height=4.cm,width=10.cm}
\caption{\small $s$-channel and $t$-channel topologies considered in the
importance sampling.}
\label{fig:6fcamp}
\ece
\efig
\ece
For the $s$-channel topology the phase-space decomposition reads:
\beqa
\nonumber
&d\Phi_6(P;q_1,\ldots ,q_6)=&(2\pi)^{12}d\Phi_2(P;Q_1,Q_2)d\Phi_2(Q_1;q_1,q_2)\\
\nonumber
&&d\Phi_2(Q_2;Q_3,Q_4)d\Phi_2(Q_3;q_3,q_4)d\Phi_2(Q_4;q_5,q_6)\\
&&dQ_1^2dQ_2^2dQ_3^2dQ_4^2,
\eeqa
where $P$ is the initial total momentum, $q_i$ are the momenta of the outgoing 
particles, while $Q_i$ are those of the internal particles. The notation
$d\Phi_2(Q_i;Q_j,Q_k)$ indicates the two-particle phase space;
the momenta of the final particles $Q_j,Q_k$
are first generated in the rest frame of the particle of momentum $Q_i$
and then boosted to the laboratory 
frame. The integration variables are the four invariant masses
$Q_1^2,\ldots ,Q_4^2$ and five pairs of angular variables $\cos\theta_i,\phi_i$,
one for each $d\Phi_2$ term. The invariant masses are sampled according to
Breit--Wigner distributions of the form
\beq
{N\over (M^2 - Q^2)^2 + \Gamma^2M^2},
\eeq
given by the propagators of the Higgs or gauge bosons in the internal lines
($N$ is a normalization factor).
For the angular variables a flat distribution is assumed. The various
$s$-channel terms differ for permutations of the external momenta and for
the parameters $\Gamma ,M$ in the importance sampling distributions.

In the case of $t$-channel diagrams, the phase space is
\beqa
\nonumber
&d\Phi_6(P;q_1,\ldots ,q_6)=&(2\pi)^{9}d\Phi_3(P;Q_1,q_1,q_2)\\
\nonumber
&&d\Phi_2(Q_1;Q_2,Q_3)d\Phi_2(Q_2;q_3,q_4)d\Phi_2(Q_3;q_5,q_6)\\
&&dQ_1^2dQ_2^2dQ_3^2,
\eeqa
where, as before, the $q_i$ are the outgoing momenta, while the $Q_i$ are
internal time-like momenta. The integration variables are the three invariant
masses, $Q_1^2,Q_2^2,Q_3^2$, three pairs of angular variables $\cos\theta,
\phi$ relative to the three $d\Phi_2$ terms, and, for the three-body phase
space $d\Phi_3$, one energy, $q_1^0$, and four angular variables,
$\cos\theta_1,\phi_1,\cos\theta_2,\phi_2$.
The invariant masses are sampled, as in the $s$-channel case, according to
Breit--Wigner distributions; one angular variable in $d\Phi_3$, say 
$\cos\theta_1$, is sampled by means of the distribution
\beq
{N\over (M_V^2 + \sqrt{s}q_1^0(1 - \cos\theta_1))^2 + \Gamma_V^2M_V^2},
\eeq
corresponding to the propagator of one space-like gauge boson ($V=W,Z$)
emitted by the initial electron or positron (typically one of the bosons
participating in the fusion into Higgs); in some channels, corresponding to
background diagrams, also another angular variable $\cos\theta$ relative to a
two-body term $d\Phi_2$, is sampled in a similar way, in order to take into
account the ``singularity'' associated with a boson propagator. All other
variables have flat distributions.

\section{Numerical results and discussion}

The numerical results presented in this section are obtained with the same
set of phenomenological parameters as adopted in ref.~\cite{sixfzpc}.
Namely, the input parameters are $G_\mu$, $M_W$ and $M_Z$, and other quantities,
such as $\sin^2\theta_W$, $\alpha$ and the widths of the $W$ and $Z$ bosons,
are computed at tree level in terms of these constants.
The Higgs width includes
the fermionic contributions $h\to \mu\mu,\tau\tau,cc,bb$, with running masses
for the quarks (to take into account QCD corrections~\cite{hwg}),
the gluonic contribution
$h\to gg$ ~\cite{hwg}, and the two-vector boson channel, according
to ref.~\cite{kniel}.
The denominators of the bosonic propagators are of the form
$p^2 - M^2 + i\Gamma M$, with fixed widths $\Gamma$.
As already discussed in ref.~\cite{sixfzpc}, the aim of this choice is to 
minimize the possible sources of gauge violation in the computation~\cite{bhf}.

Such gauge violations have been studied by the same
methods as were used in ref.~\cite{sixfzpc}. In particular, for what concerns $SU(2)$
gauge symmetry, comparisons have been made with results in the so-called
``fudge scheme''~\cite{fudge}.
A disagreement has been found at the level of few per cent
for a c.m. energy of $360$ GeV. The disagreement vanishes at higher energies.
By careful inspection of the various contributions, it has been checked 
that the deviation at lower energies is due to the well-known fact that a given
fudge factor, close to a resonance, mistreats the contributions that do not
resonate in the same channel.
Concerning $U(1)$ invariance, a test has been performed by using different
forms of the photon propagator and finding perfect agreement,
up to numerical precision, in the values
of the squared matrix element.

The first group of results discussed in this section refers to cross-section
calculations, performed by using the program as an integrator of weighted
events. The signature considered in the first plots of total cross section
is $q\overline ql^+l^-\nu\overline\nu$, where, in addition to the sums over
quark and neutrinos flavours already mentioned, there is a sum over
$l=e,\mu,\tau$. All other results are instead relative to the signature
$q\overline qe^+e^-\nu\overline\nu$.
Some samples of unweighted events, obtained by using the code as a generator,
are then analysed in the remaining part of this section.

\subsection{Total cross sections}

In fig.~\ref{fig:6fsscan} the total cross section (including the contribution
of all the tree-level Feynman diagrams) is shown for three values of the Higgs
mass in the intermediate range,
$165,185$ and $255$ GeV, at energies between $360$ and $800$ GeV. To make
a first analysis, the following kinematical cuts are adopted: the invariant
mass of the quark--antiquark pair and that of the charged lepton pair are
 required to be
greater than $70$ GeV, the angles of the two charged leptons with respect to
the beam axis within $5^\circ$ and $175^\circ$.
This choice
is applied to the quantities shown in figs.~\ref{fig:6fsscan},
\ref{fig:6f185isr} and~\ref{fig:6ft-spbg}.
Another set of cuts, with a lower limit of $20$ GeV
on the $l^+l^-$ invariant mass, is adopted in fig.~\ref{fig:6fsscan20} and
in the study of event samples.

The increase with energy, common to all three curves in
fig.~\ref{fig:6fsscan}, is due, at high energies, to the $t$-channel
contributions; in the case of $m_H=255$ GeV, the steep rise near
$\sqrt{s}=360$ GeV is related to the existence of a threshold
effect for the Higgs-strahlung process at an energy $\sqrt{s}\sim m_H + M_Z$.

\bce
\bfig
\bce
\epsfig{file=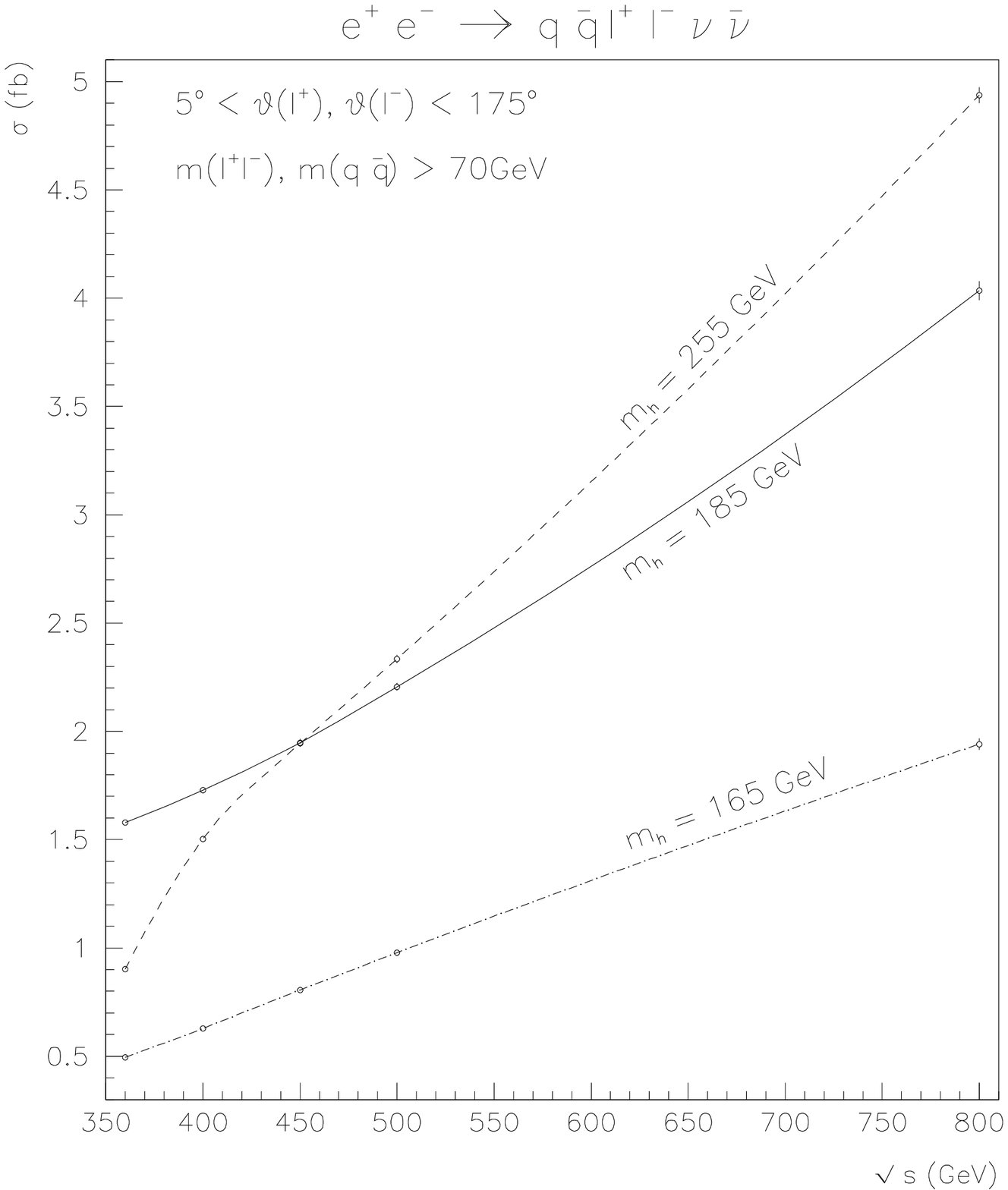,height=8.cm,width=8.cm}
\caption{\small Total cross section for the process 
 $e^+e^-\to q\overline ql^+l^-\nu\overline\nu$ in the Born approximation,
 as a function of $\sqrt s$ for three different values of the Higgs mass
$m_H$. The angles $\theta(l^+)$, $\theta(l^-)$
of the charged leptons with the beam axis are in the interval 
$5^\circ$-$175^\circ$,
the $e^+e^-$ and the $q\bar q$ invariant masses are larger than $70$~GeV.}
\label{fig:6fsscan}
\ece
\efig
\ece

In fig.~\ref{fig:6fsscan20} the total cross section is plotted, with the cut on
the invariant mass of the charged lepton pair reduced to $20$ GeV. The effect
of this modification, as expected, is an enhancement of the cross section in
the low-energy region: indeed the most important contribution at energies below
$500$ GeV is given by the Higgs-strahlung diagram, with the Higgs decaying into
two $W$ bosons, which will be indicated from now on as the charged-current
Higgs-strahlung diagram, and which is characterized by a broad distribution of
the $l^+l^-$ invariant mass that goes well below $70$ GeV. This cut is still
sufficient to reduce to a negligible size the contribution of virtual photon
conversion into $l^+l^-$ pairs. The behaviour of the cross section as the Higgs
mass is varied depends on the interplay of the various production mechanisms
and of the decay branchings involved; this behaviour can be better observed in
the ``signal'' contribution that will be defined below
(see fig.~\ref{fig:sigmh}).
\bce
\bfig
\bce
\epsfig{file=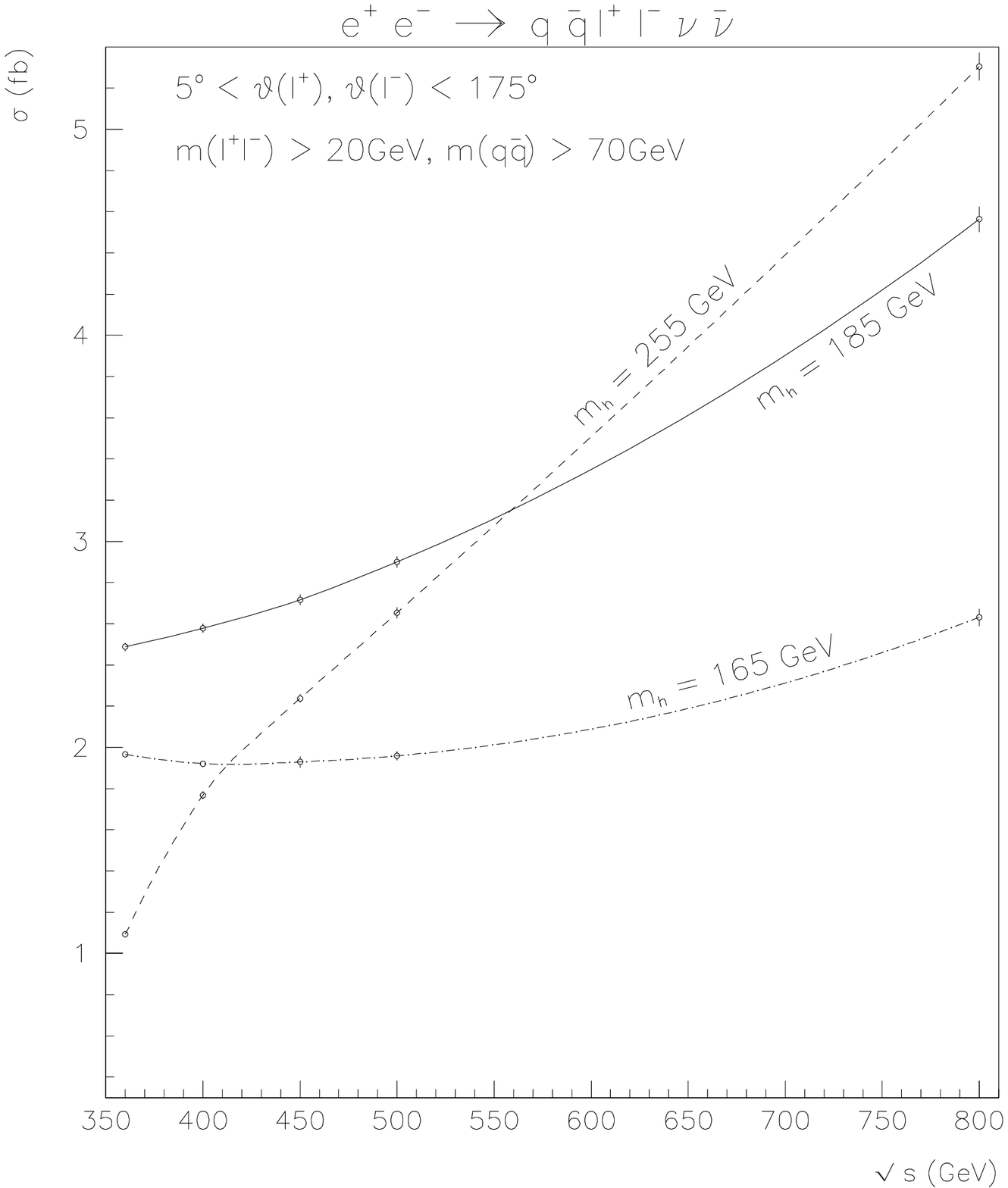,height=8.cm,width=8.cm}
\caption{\small The same as in fig.~\ref{fig:6fsscan} with the cut on
the $l^+l^-$ invariant mass reduced to $20$ GeV.}
\label{fig:6fsscan20}
\ece
\efig
\ece

The effect of ISR is illustrated in fig.~\ref{fig:6f185isr}, for a Higgs mass of
$185$ GeV and for the signature $q\overline qe^+e^-\nu\overline\nu$, to which
all the remaining results refer. Here the $e^+e^-$ invariant mass is again
greater than $70$ GeV.

\bce
\bfig
\bce
\epsfig{file=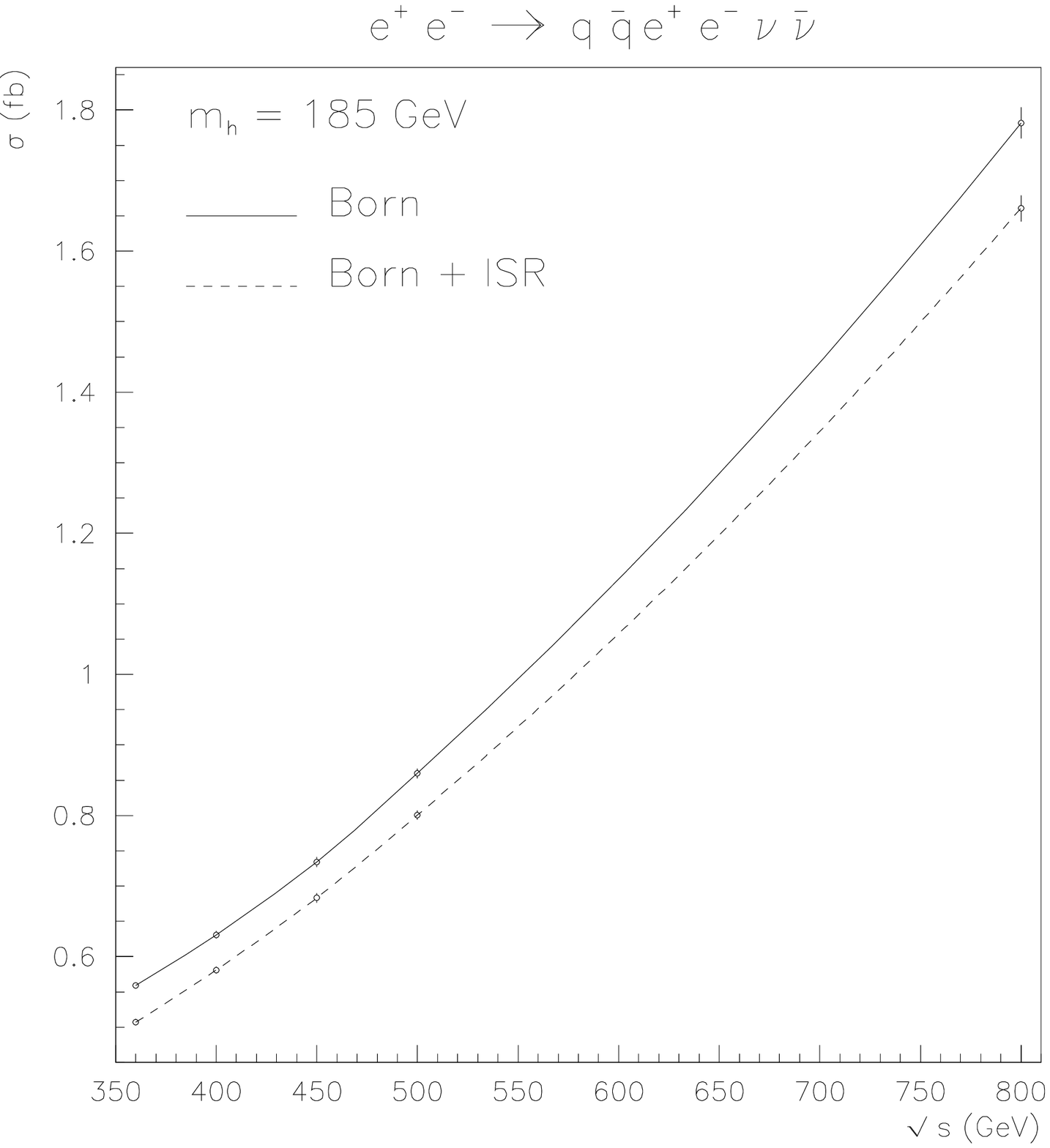,height=7.5cm,width=7.5cm}
\caption{\small Effect of initial-state radiation on the total cross section
of the process $e^+e^-\to q\overline qe^+e^-\nu\overline\nu$ as a function
of $\sqrt s$ for a Higgs mass $m_H=185$ GeV. Cuts are the same as in 
fig.~\ref{fig:6fsscan}.}
\label{fig:6f185isr}
\ece
\efig
\ece

The cross section in the presence of ISR is lowered by a quantity of the order 
of $10\%$ with respect to the Born approximation. This phenomenon can be
easily understood, since the initial-state radiation reduces the c.m. energy,
so in this case it produces a shift towards energy values where the cross
section is smaller.

\subsection{Definition and study of the Higgs signal}

The results discussed so far, as stated above, are given by the sum of all the
tree-level Feynman diagrams. Strictly speaking, this is the only meaningful
procedure. On the other hand, there is a number of reasons to
consider a subset of diagrams that can be defined as the Higgs signal and to
define a corresponding background. In the first place this is of great interest
from the point of view of the search for the Higgs boson in the experiments.
Moreover, as will be shown, such a definition allows one to make a comparison 
with results obtained in the narrow width approximation 
(NWA)~\cite{hstr}--\cite{zervast},
which are the only available
estimations unless a complete $6f$ calculation is performed.
In principle, whenever a subset of diagrams is singled out, gauge invariance
may be lost and unitarity problems may arise. However, in the following, an
operative definition of signal and background is considered and its reliability
is studied for various Higgs masses and c.m. energies.

The signal is defined as the sum of the six graphs containing a resonant Higgs
boson, shown in fig.~\ref{fig:6fd}. The background is defined as the sum of all
the diagrams without a Higgs boson. In this definition the diagrams with a 
non-resonant Higgs boson coupled to gauge bosons, shown in fig.~\ref{fig:6fdht}, 
are missing both in the signal and in the background. Such a choice has been
dictated by the fact that these non-resonant contributions cannot correctly be
included in the signal, since they cannot find a counterpart in the NWA, and
because of gauge cancellations with background
contributions at high energies; however, as they depend on the Higgs mass, 
they should not be included in the background as well.

In order to give a quantitative estimate of the validity of this definition,
the total cross section (sum of all the tree-level
$6f$ Feynman diagrams) is compared in fig.~\ref{fig:6ft-spbg} with the 
incoherent sum of signal and background. 
The cuts are as in fig.~\ref{fig:6fsscan}, in particular
with the $e^+e^-$ invariant mass greater than $70$ GeV. In order to understand
the meaning of these results, it is important to note that, as observed above,
the contributions of the diagrams of fig.~\ref{fig:6fdht} are absent both from
the signal and from the background: thus if we indicate these contributions to
the scattering amplitude as $A_{ht}$, and the signal and background amplitudes
as $A_s$ and $A_b$ respectively, the total squared amplitude is
\beq
\vert A\vert^2 =\vert A_s + A_b + A_{ht}\vert^2 \quad .
\eeq
The terms neglected in the incoherent sum of signal and background are
$\vert A_{ht}\vert^2$ and all the interference terms. Among these, the
interferences of $A_{ht}$ with the rest are dominant at high energies
as they involve gauge cancellations.

\bce
\bfig
\bce
\epsfig{file=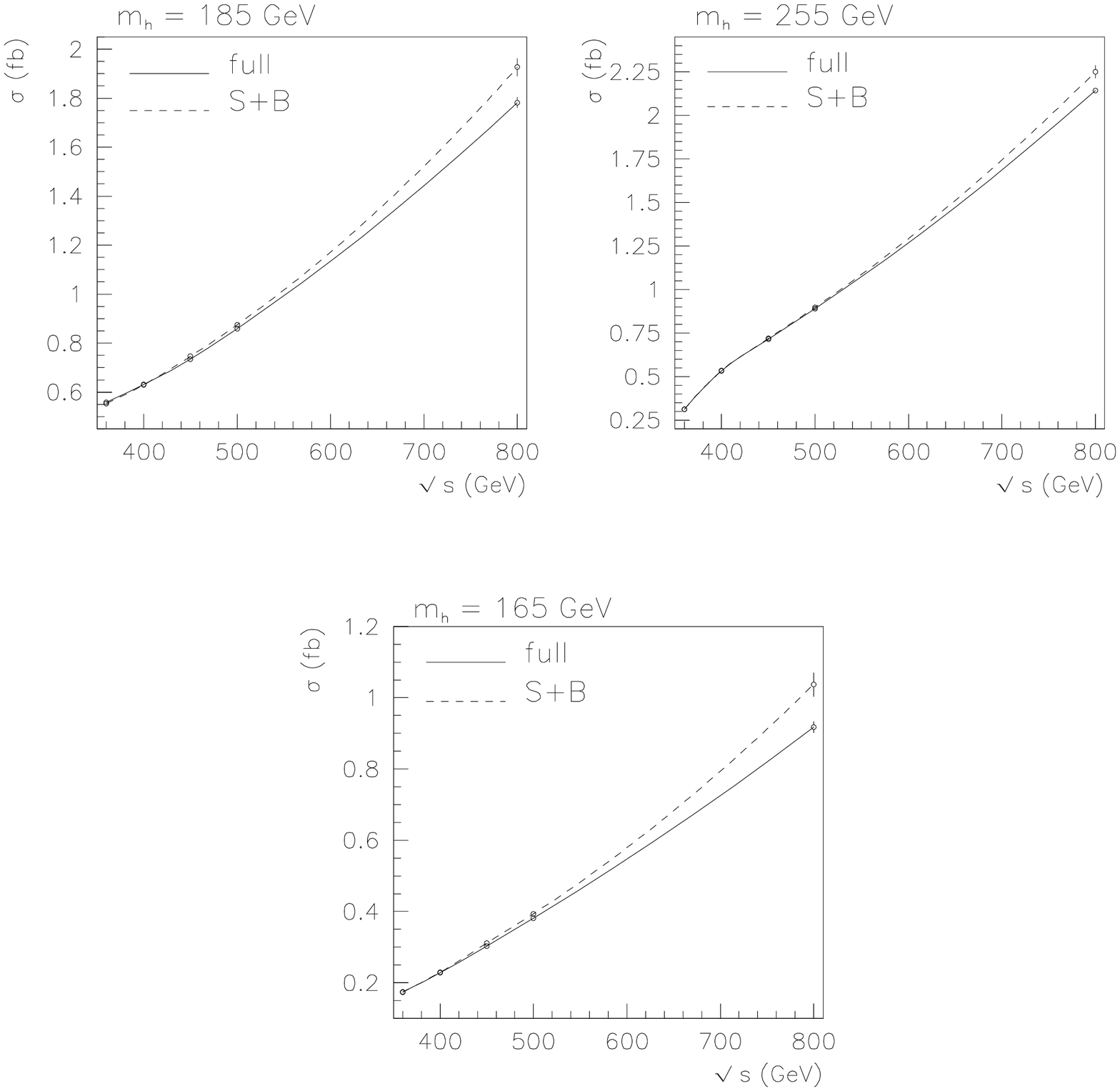,height=11.cm,width=11.cm}
\caption{\small Full six-fermion cross section compared with the incoherent sum
of ``signal'' (S) and ``background'' (B).
A detailed discussion and an operative definition of ``signal''  
and ``background'' are given in the text.}
\label{fig:6ft-spbg}
\ece
\efig
\ece

The curves of fig.~\ref{fig:6ft-spbg} show that up to $500$ GeV the total cross
section and the incoherent sum of signal and background are indistinguishable
at a level of accuracy of $1\%$,
and the definition of signal may be considered meaningful; at higher
energies, this separation of signal and background starts to be less reliable,
since it requires to neglect effects that are relevant at this accuracy. In
particular, at $800$ GeV the deviation is of the order of a few per cent and it
decreases when the Higgs mass passes from $165$ to $185$ and to $255$ GeV.

The above results are obtained with the set of kinematical cuts in which the
$e^+e^-$ invariant mass is greater than $70$ GeV, but when this cut is at
$20$ GeV, the difference between full cross section and incoherent sum of
signal and background is significantly reduced (about $3$--$4\%$ at $800$ GeV).
The analysis of event samples
presented in the following is made within this latter set of cuts, so that, up
to 800 GeV, it can be considered reliable, at the level of accuracy of a few
per cent, to speak of ``background'', as will be done.

On the other hand the problems arising when a definition of signal and
background is attempted show the importance of a calculation involving the full
set of tree-level Feynman diagrams to obtain reliable results, especially at
high energies.

A comparison with the NWA is shown in 
fig.~\ref{fig:6ff-nwa} for the processes
$e^+e^-\to q\overline q e^+ e^- \nu\overline\nu$ and
$e^+e^-\to q\overline q \mu^+ \mu^- \nu\overline\nu$, where, for the sake of
comparison, no kinematical cuts are applied and the results are in the Born
approximation. Here $\sigma_{sig}$
is the signal cross section, containing the contributions of the six diagrams of
fig.~\ref{fig:6fd} (or the suitable subset of these for the case of the final
state $q\overline q \mu^+ \mu^- \nu\overline\nu$) and their interferences.
The cross section in the NWA, $\sigma_{NWA}$, is obtained in the following way
(for definiteness the case with $e^+e^-$ in the final state is considered): the
known cross sections for the processes of real Higgs production 
$e^+e^-\to h\nu\overline\nu,he^+e^-$ ~\cite{almele,zervast} and 
$e^+e^-\to Zh$~\cite{hstr} are multiplied by the appropriate branching ratios,
so as to obtain six terms corresponding to the diagrams of fig.~\ref{fig:6fd};
then the incoherent sum of these terms is taken.
Thus the comparison between $\sigma_{sig}$ and $\sigma_{NWA}$
gives a measure of interference between the different production mechanisms
and of off-shellness effects together.
\bce
\bfig
\bce
\epsfig{file=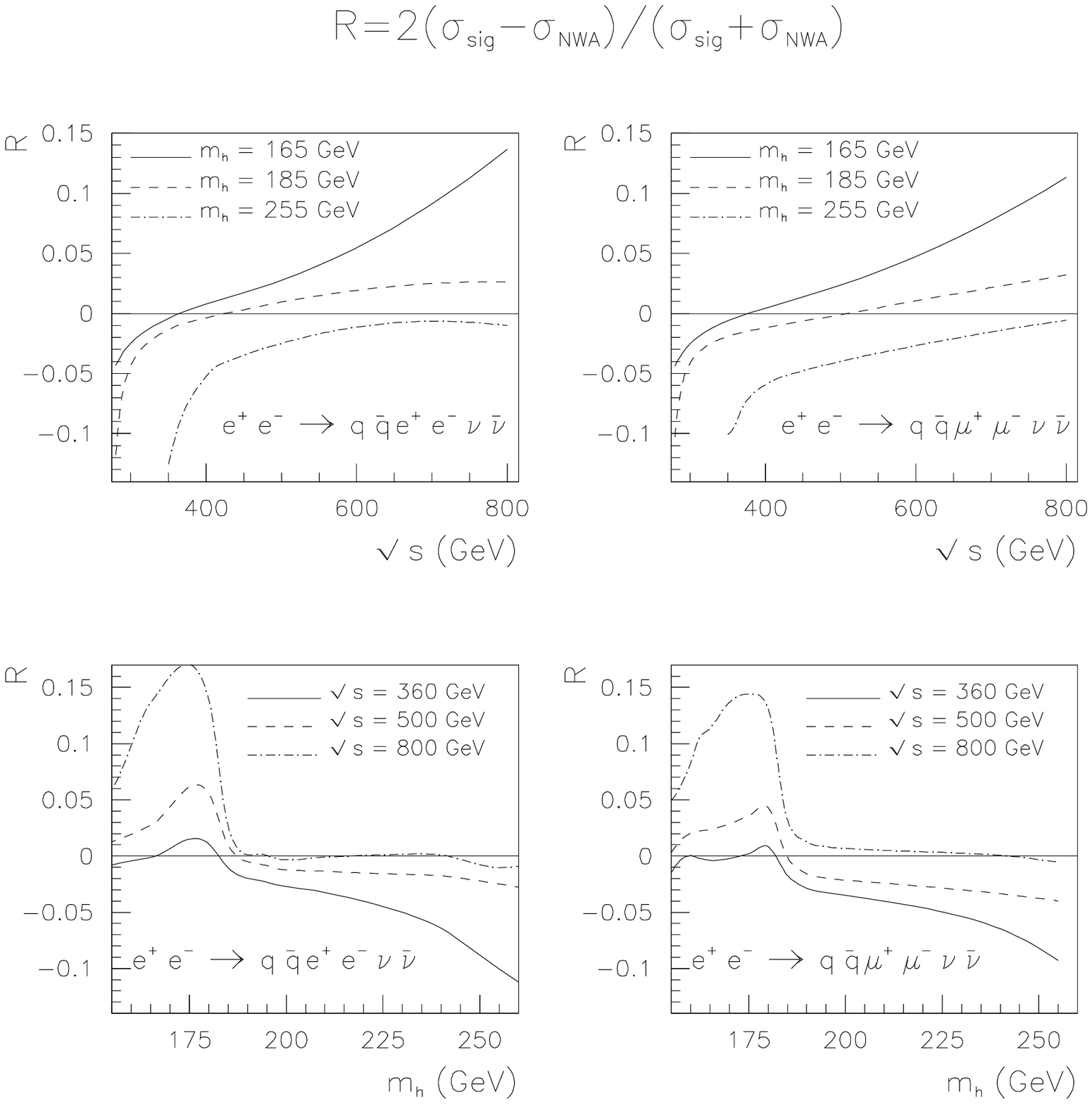,height=14.cm,width=14.cm}
\caption{\small Comparison between the signal cross section obtained by a
diagrammatic six-fermion calculation and the one calculated in the
narrow width approximation (see the discussion in the text),
as a function of $\sqrt s$ (upper row) and of the Higgs mass
(lower row).}
\label{fig:6ff-nwa}
\ece
\efig
\ece
As can be seen in fig.~\ref{fig:6ff-nwa}, the relative difference $R$ is of the
order of
some per cent, depending on the Higgs mass and the c.m. energy; in some cases
it reaches values of more than $10\%$, with no substantial difference between
the two final states considered. In particular the off-shellness effects
are much more important than the interference ones. In fact the relative size
of the interferences has been separately evaluated by means of a comparison
between $\sigma_{sig}$ and the incoherent sum of the six diagrams of
fig.~\ref{fig:6fd} and has been found to be at most $2\%$, but generally less
than $1\%$ for the c.m. energies and Higgs masses considered here.\par
The size of the off-shellness effects, comparable with the ISR lowering,
indicates the importance of a full
$6f$ calculation in order to obtain sensible phenomenological
predictions.

In fig.~\ref{fig:sigmh} the signal cross section is shown as a function of the
Higgs mass for different c.m. energies.
The behaviour is related to the branching ratios of the decays of
the Higgs boson into gauge bosons and the differences between the three energy
values considered are due to the variations in the relative sizes of the
different signal contributions at different energies.
\bce
\bfig
\bce
\epsfig{file=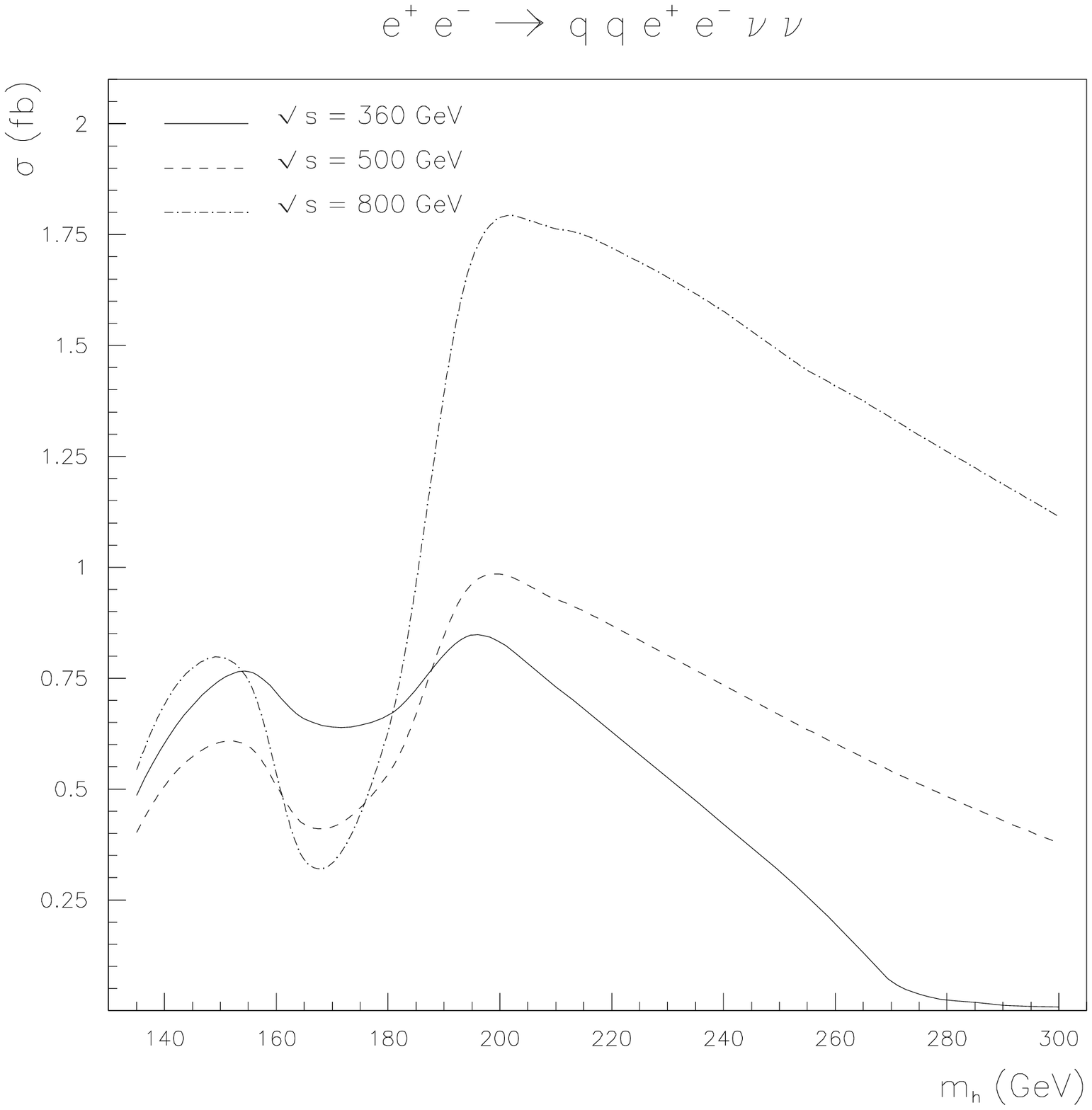,height=8.cm,width=8.cm}
\caption{\small Signal cross section as a function of the Higgs mass for three
different c.m. energies.}
\label{fig:sigmh}
\ece
\efig
\ece

\vskip 5 truemm

\subsection{Distributions}

The results presented in the following refer to samples of unweighted events
for a Higgs mass of $185$ GeV at energies of $360$ and $800$ GeV, with or
without the effect of ISR and beamstrahlung. The cuts adopted in all cases are
the following: the invariant mass of the $q\overline q$ pair greater than $70$
GeV, the invariant mass of the $e^+e^-$ pair greater than $20$ GeV, and the
angles of the electron and positron with respect to the beam axis between
$5^\circ$ and $175^\circ$; further cuts, applied in the analysis of particular
cases, will be described later. The numbers of events in all the samples are
normalized to the same luminosity.

In fig.~\ref{fig:nt185ibs4} the invariant masses of two different systems of
four momenta are studied at c.m. energies of $360$ and $800$ GeV in the Born
approximation (dashed histograms) as well as with ISR and beamstrahlung
(solid histograms). The first set is given by
$e^+e^-+$ missing momentum, where the missing momentum is defined as
$q_{miss}=p^+_{in}+p^-_{in}-q_{e^+}-q_{e^-}-q_q-q_{\overline q}$.
In the Born approximation this set of momenta corresponds to the system
$e^+e^-\nu\overline\nu$. The other set considered is that corresponding to the
four-fermion system $q\overline qe^+e^-$.
As can be seen in fig.~\ref{fig:6fd}, these are two of the possible sets of 
four fermions produced by the decay of the Higgs boson in the process under
consideration; there is also a third set, $q\overline q\nu\overline\nu$, whose
invariant mass distribution, however, does not contain any new feature.
The presence of the Higgs boson can be revealed by a peak in the 
distributions of these invariant masses. Indeed, in the Born approximation 
(dashed histograms), a sharp peak around $185$ GeV can be seen in each of the
histograms of fig.~\ref{fig:nt185ibs4}. At a c.m. energy of $360$ GeV, the most
remarkable one is that of $e^+e^-+$ missing momentum, associated
to the system $e^+e^-\nu\overline\nu$, as it receives contributions from the
charged current Higgs-strahlung diagram, which is dominant at this energy.
In the presence of ISR and beamstrahlung, this peak is considerably lowered and
broadened, while the other distribution, not involving the missing momentum,
is not significantly affected by radiative effects.
At $800$ GeV this phenomenon is even more evident, because the peak in the first
distribution is completely eliminated by radiative effects, as a consequence of
the small size of the charged current Higgs-strahlung term at this energy,
while the second distribution results to be very sensitive to the presence of
the Higgs, since it receives, around $185$ GeV, contributions from the diagram of
$WW$ fusion into Higgs, which is the dominant signal term at high energies, and
the presence of ISR and beamstrahlung does not modify the shape of the
histogram. Thus, at high energies, a very clean signal of the
Higgs boson is provided by the process under study through this distribution.

\bce
\bfig
\bce
\epsfig{file=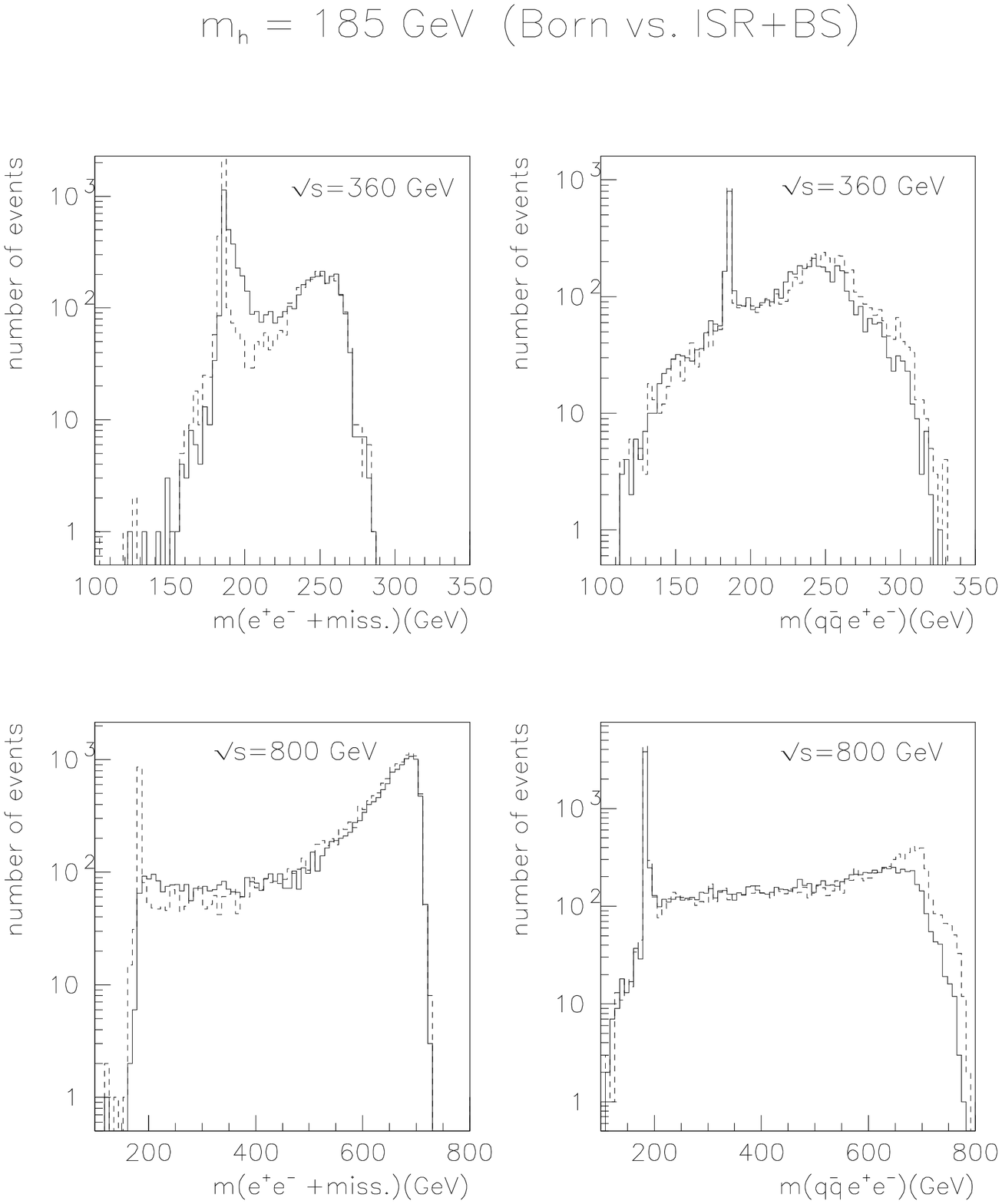,height=11.cm,width=11.cm}
\caption{\small Invariant-mass distributions for four-fermion systems in the
Born approximation (dashed histograms) and with ISR and beamstrahlung (solid
histograms) at $\sqrt s=360$ GeV (upper row) and $\sqrt s=800$ GeV (lower row).}
\label{fig:nt185ibs4}
\ece
\efig
\ece

The quantities analysed above are useful to reveal the presence of the Higgs
boson and to determine its mass. Other variables can be considered to study
the properties of this particle, such as spin and parity. Some examples are
considered in figures~\ref{fig:th12b},~\ref{fig:cthzz} and~\ref{fig:thzw0}.

When the process $e^+e^-\to HZ$ is considered, a variable that can give
evidence of the scalar nature of the Higgs is the angle $\theta_Z$ of the $Z$
particle direction with respect to the beam in the laboratory frame. 
It is well known~\cite{hstr} that the differential cross section 
$d\sigma/d\cos\theta_Z$
goes as $\sin^2\theta_Z$
at energies much greater than $M_Z$ and away from the threshold for Higgs
production. A similar situation is expected to occur for the $6f$
process under study when the Higgs-strahlung contributions are dominant.
The distribution $d\sigma/d\theta_Z$ is shown, at the c.m. energies of $360$
and $800$ GeV,
in fig.~\ref{fig:th12b}, where the $Z$ particle is reconstructed as
the sum of the quark and antiquark momenta (indeed this is the case for the
dominant diagram). The contribution from the background alone (dashed histogram)
is also shown. The shape of the solid histogram shows
the expected behaviour at $360$ GeV, where Higgs-strahlung dominates.

\bce
\bfig
\bce
\epsfig{file=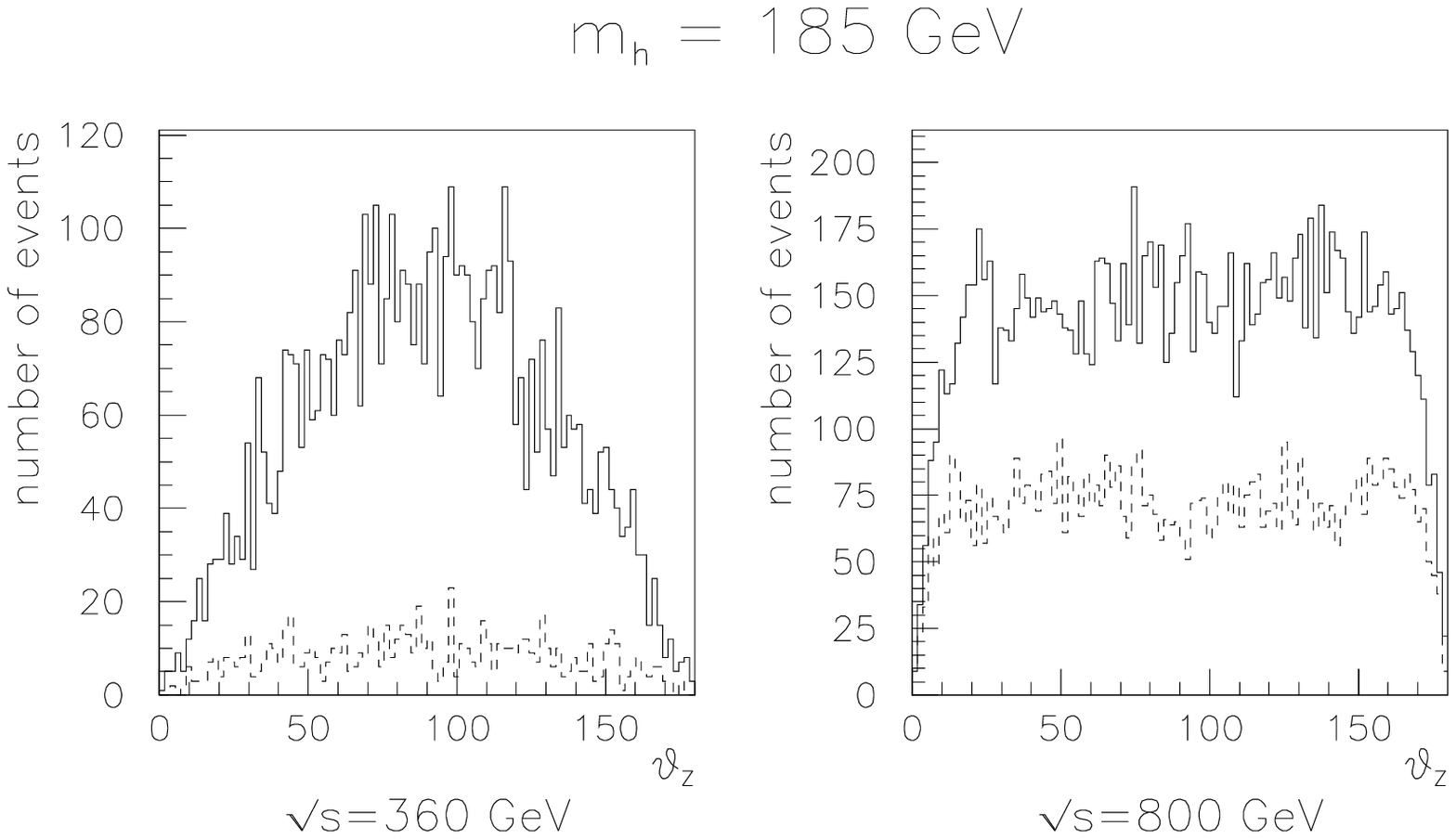,height=8.cm,width=14.cm}
\caption{\small Distribution of the angle $\theta_Z$
of the $q\bar q$ pair with respect to the beam in the
laboratory frame at $\sqrt s=360,800$ GeV.
The solid histogram represents the full calculation, the dashed histogram is
the contribution of the background.}
\label{fig:th12b}
\ece
\efig
\ece

At the c.m. energy of $800$ GeV, where the dominant signal diagram is
$WW$ fusion into Higgs, the situation is substantially different, since the
process of Higgs production is of the $t$-channel type. One variable that results
very sensitive to the presence of the Higgs boson is shown in
fig.~\ref{fig:cthzz} and indicated as $\cos\theta_{ZZ}$; $\theta_{ZZ}$ is the
angle between the three-momenta in the laboratory frame
of the $q\bar q$ and $e^+e^-$ pairs, which
correspond, in the diagram of $WW$ fusion, to the $Z$ particles coming from the
Higgs. The full distribution (solid line) and the contribution from the
background alone (dashed line) are particularly distinguished in the region
near $1$. There is here a clear signal of the presence of the Higgs, and such a
variable can be used to impose kinematical cuts to single out signal
contributions. This phenomenon is however of a kinematical nature, and is not
directly related to the scalar nature of the Higgs, but is rather a
consequence of the smallness of diagrams with the same topology of the $WW$
fusion in the background.

\bce
\bfig
\bce
\epsfig{file=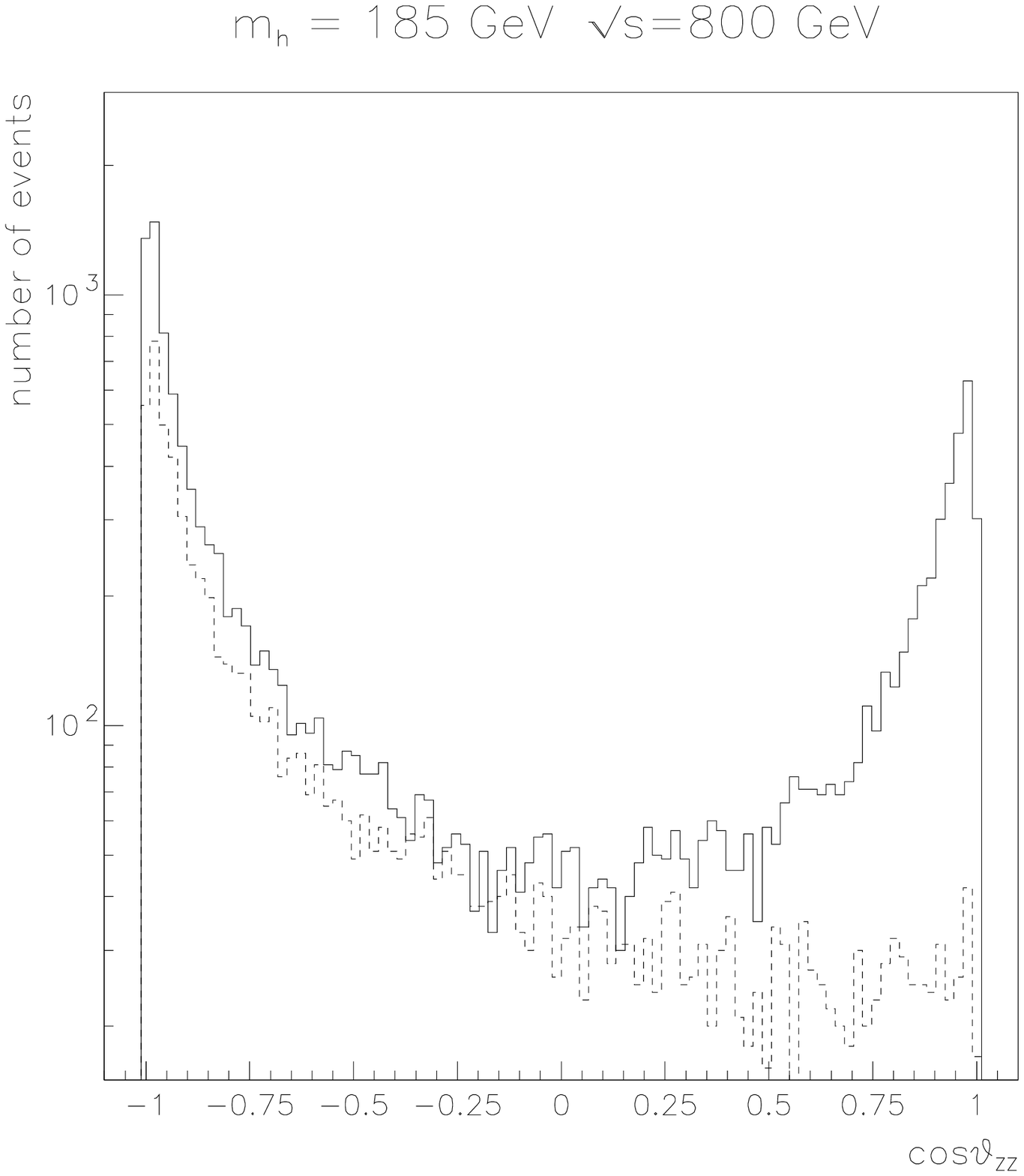,height=8.cm,width=8.cm}
\caption{\small Distribution of the angle $\theta_{ZZ}$
between the $q\bar q$ and the $e^+e^-$ pairs in the
laboratory frame at $\sqrt s=800$ GeV. The solid histogram is the full
calculation, the dashed histogram is the background.}
\label{fig:cthzz}
\ece
\efig
\ece

Another variable has been considered at the energy of $800$ GeV in
fig.~\ref{fig:thzw0}. It is the angle $\theta^*_Z$ of the $Z$ particle,
reconstructed as the sum of the electron and positron momenta, with respect to
the beam, in the rest frame of the system $q\bar qe^+e^-$. The reference diagram
is always the $WW$ fusion: this may be regarded asymptotically as an $s$-channel
$WW$ scattering into $ZZ$, and, in the rest frame of the incoming $WW$ pair,
the angular distribution of the produced $ZZ$ pair is determined by the
scalar nature of the exchanged particle.
In the first row of fig.~\ref{fig:thzw0} the plot on
the left is made without additional cuts, while the plot on the right is
obtained with the requirement $\cos\theta_{ZZ}>0$, so as to reduce the
background. In the 
second row the invariant mass of $q\bar qe^+e^-$ is required
to be smaller than 250 GeV (left) and
within 20 GeV around the Higgs mass (right) in order to further suppress the
background.
A clear difference between the shape of the full distribution and
that of the background can in fact be seen, and in the last three plots the
behaviour is very similar to the $\sin\theta^*_Z$ distribution expected on the
basis of the above observations.

\bce
\bfig
\bce
\epsfig{file=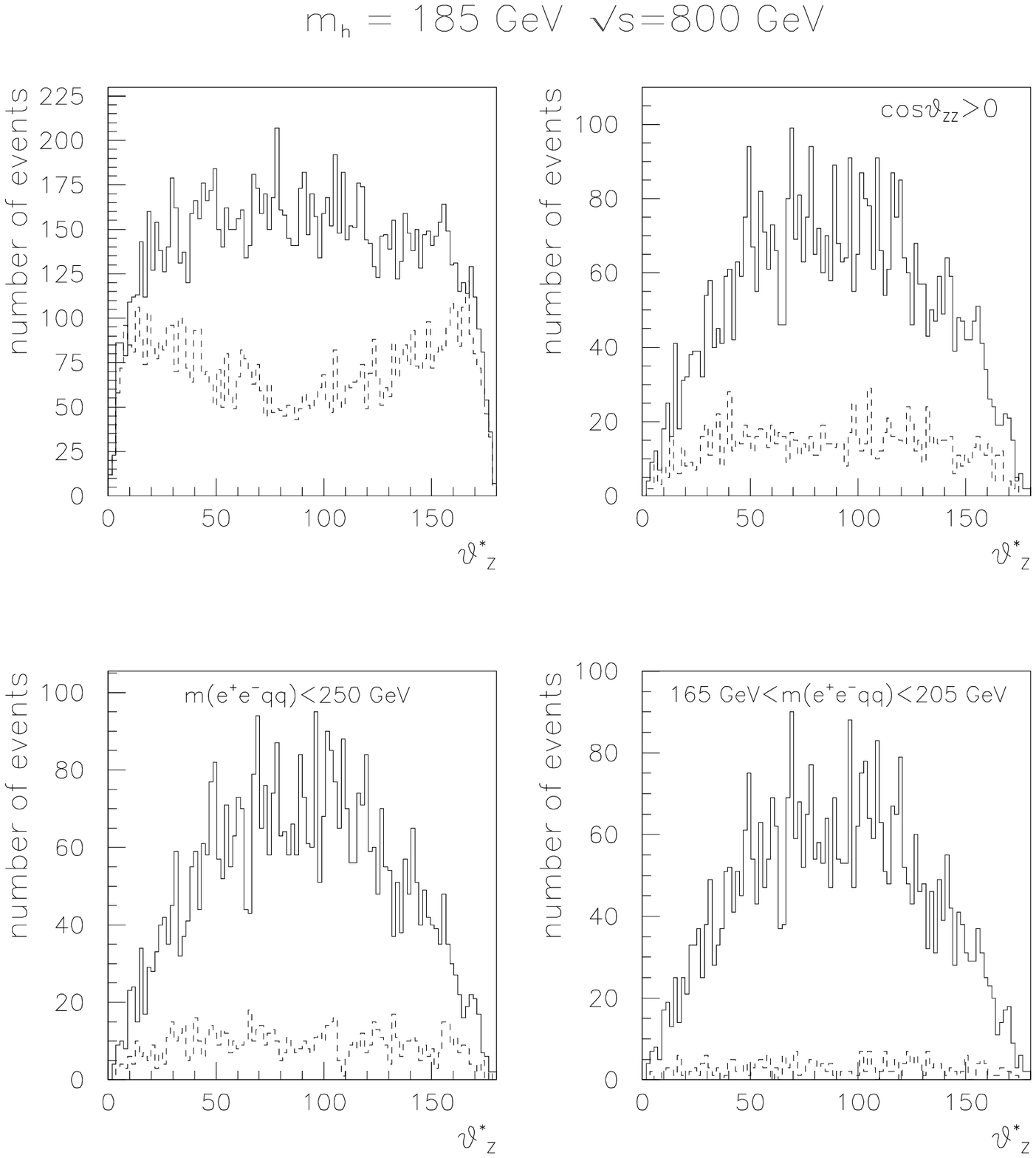,height=14.cm,width=14.cm}
\caption{\small Distribution of the angle $\theta^*_{Z}$
of the $e^+e^-$ pair with respect to the beam axis in the
rest frame of the $q\bar qe^+e^-$ system at $\sqrt s=800$ GeV. The solid
histogram is the full calculation, the dashed histogram is the
background. First row: no additional cuts (left), $\cos\theta_{ZZ}>0$ (right),
where the angle $\theta_{ZZ}$ is defined in the text and shown in
fig.~\ref{fig:cthzz}. Second row: $q\bar qe^+e^-$ invariant mass smaller than
250 GeV (left) and within 20 GeV around the Higgs mass (right).}
\label{fig:thzw0}
\ece
\efig
\ece

\section{Conclusions}

The processes $e^+e^-\to q\overline q l^+ l^- \nu\overline\nu$ have been studied
in connection with the search for an intermediate-mass Higgs boson. The study,
which extends a previous analysis of $6f$ signatures with only two jets,
is characterized by the presence of neutral current contributions that were 
never considered before and by the fact that several mechanisms 
of Higgs production are simultaneously active.

The tool used for the numerical calculations is a Fortran code based on the
algorithm ALPHA, for the determination of the scattering amplitude, and on a
development of the Monte Carlo program HIGGSPV/WWGENPV, for the phase-space
integration.

The total cross section, including all the tree-level Feynman diagrams, has
been calculated with various kinematical cuts and taking into account the
effects of ISR and beamstrahlung.

A definition of signal and background has been considered and its reliability
has been studied. To this end the incoherent sum of ``signal'' and
``background'' has
been compared with the full cross section, and this has shown deviations that,
up to a c.m. energy of $500$ GeV, are negligible to an accuracy of $1\%$, but may be
of several per cent at $800$ GeV (fig.~\ref{fig:6ft-spbg}).
These deviations are, however, reduced when the
kinematical selection criteria become more inclusive. 

A comparison of the ``signal'' cross section with results in the NWA
has shown that off-shellness effects have a relative size of
several per cent (fig.~\ref{fig:6ff-nwa}).

The results of figs.~\ref{fig:6ft-spbg} and \ref{fig:6ff-nwa} show the
importance of a complete $6f$ calculation to produce reliable results
at the TeV scale.

In the study of generated events the problem of finding observables that are 
sensitive to the presence of the Higgs and to its properties has been 
addressed. The
presence of several mechanisms of Higgs production, whose relative importance
varies with energy, requires that different variables be considered according 
to the energy range studied.

The invariant masses of two sets of four fermions have been analysed first
(fig.~\ref{fig:nt185ibs4}): one, relative to the system $e^+e^-+$missing
momentum, is relevant to the detection of the Higgs boson at $360$ GeV of c.m.
energy, but, at $800$ GeV, the effects of ISR and beamstrahlung
prevent to study the Higgs by means of this distribution.
The other invariant mass, relative to the system $e^+e^- q\bar q$, is instead
particularly useful at high energies and is almost completely unaffected by
radiative effects.

Three angular variables have then been studied: the angle $\theta_Z$
(fig.~\ref{fig:th12b}) is suited
to reveal the spin zero nature of the Higgs at $360$ GeV, where the
Higgs-strahlung dominates, but it gives no information at $800$ GeV.
The angles $\theta_{ZZ}$ (fig.~\ref{fig:cthzz}) and $\theta^*_Z$
(fig.~\ref{fig:thzw0}) are very useful at $800$ GeV: the first one is very
effective to single out the signal, but is not able to distinguish the spin
nature of the Higgs; the second one has a distribution whose shape is very
different from that of the background, and is related to the spinless nature
of the Higgs particle.

The computing strategy and the relative computer code
developed in this work have been applied to study
intermediate-mass Higgs physics.
However, the variety
of diagram topologies present in the matrix element and
taken into account in the Monte Carlo integration,
as well as the possibility provided by ALPHA of dealing with any
kind of process, including now also QCD amplitudes, give the opportunity to
examine other topics relevant to physics at future colliders, where
$6f$ production is involved.

\vspace{1.truecm}
\noindent
{\bf Acknowledgements}\\
We wish to thank A.~Ballestrero and T.~Ohl for discussions and
for their interest in our work. The work of M.~Moretti is funded 
by a Marie Curie fellowship (TMR-ERBFMBICT 971934). 
F.~Gangemi thanks the INFN, Sezione di Pavia, for the use of 
computing facilities.

\end{document}